\documentclass[10point, reprint, twocolumn, aps, prl]{revtex4}
\usepackage{graphicx}
\usepackage{color,soul}
\usepackage{float}
\usepackage{amsmath}
\usepackage{bm}
\usepackage{amssymb}
\usepackage{amsfonts}
\usepackage{dcolumn}
\usepackage{epsfig}
\usepackage{subfigure}
\usepackage{bm}
\begin{document}

\title{Magnetothermopower of $\delta$-doped LaTiO$_3$/SrTiO$_3$ interfaces in the Kondo regime}

\author{Shubhankar Das$^{1}$, P. C. Joshi$^{1}$, A. Rastogi$^{1}$, Z. Hossain$^{1}$ and R. C. Budhani$^{1,2,}$}
\email{rcb@iitk.ac.in, rcb@nplindia.org}
\affiliation{$^1$Condensed Matter - Low Dimensional Systems Laboratory, Department of Physics, Indian Institute of Technology, Kanpur 208016, India\\
$^2$National Physical Laboratory, Council of Scientific and Industrial Research (CSIR), New Delhi - 110012, India}

\date{\today}
\begin{abstract}
Measurements of magneto-thermopower (S(H, T)) of interfacial delta doped LaTiO$_3$/SrTiO$_3$ (LTO/STO) heterostructure by an iso-structural antiferromagnetic perovskite LaCrO$_3$ are reported. The thermoelectric power of the pure LTO/STO interface at 300 K is $\approx$ 118 $\mu$V/K, but increases dramatically on $\delta$-doping. The observed linear temperature dependence of S(T) over the temperature range 100 K to 300 K is in agreement with the theory of diffusion thermopower of a two-dimensional electron gas. The S(T) displays a distinct enhancement in the temperature range (T $<$ 100 K) where the sheet resistance shows a Kondo-type minimum. We attributed this maximum in S(T) to Kondo scattering of conduction electron by localized impurity spins at the interface. The suppression of S by a magnetic field, and the isotropic nature of the suppression in out-of-plane and in-plane field geometries further strengthen the Kondo model based interpretation of S(H, T).

\end{abstract}
\maketitle


The interface formed at the surface of TiO$_2$ layer terminated SrTiO$_3$ (STO) crystals and a few unit cells (uc) thick epitaxial overlayers of LaTiO$_3$ (LTO) and LaAlO$_3$ (LAO) has been established to show exotic electronic phases that include superconductivity and ferromagnetism \cite{Ohtomo1, Ohtomo2, Reyren, Li, Dikin, Brinkman, Rastogi, Biscaras1}. The electronic transport through the interfacial two-dimensional electron gas (2DEG), even in the absence of such collective phenomena, has novel features driven by anisotropy, a broken inversion symmetry (BIS) and localized moments at Ti$^{3+}$ ion sites \cite{Shalom, Caviglia, Lerer}. The BIS leads to a strong Rashba spin-orbit (S-O) interaction \cite {Bychkov&Dresselhaus} and the attendant weak antilocalization (WAL) \cite{Bergmann}, where as the Ti$^{3+}$ ions with 3d$^1$ configuration create localized S = 1/2 spins that lead to s-d scattering, conduction electron mediated coupling of spins and even a robust Kondo-like \cite{Kondo1} higher order scattering \cite{Brinkman, Shalom, Shubhankar}. The strength of some of these features, however, depends strongly on the conditions employed for growth of LAO and LTO overlayers. Films grown in oxygen rich environment (pO$_2$ $>$ 10$^{-4}$ mbar) tend to have a less steeper metallic sheet resistance (R$_\Box$(T)), a well defined minimum (T$_m$) in R$_\Box$ in the temperature range of 10 to 50 K, and a ln T growth of R$_\Box$ at T $<$ T$_m$ followed by saturation at still lower temperatures \cite{Kalabukhov, Wang}. These films also show a large positive magnetoresistance (MR) \cite{Lerer}. Such features in R$_\Box$(T) are ubiquitous signatures of Kondo scattering in noble metals like Cu and Au doped with small amounts of 3d transition metals such as Mn, Fe or Co \cite{Chen, Blachy&Borda, Blachy&Tersoff}. Indeed, the Kondo scattering was agreed to be the source of the resistivity minimum seen in LAO/STO interface by Brinkman et al. \cite{Brinkman}, although they did not pinpoint the magnetic scatterer. A Kondo like resistivity has also been seen by Lee et al. \cite{Lee} in the 2DEG induced at STO surface by electrolytic gating. These  authors have argued that the localized Ti$^{3+}$ d-electron, acting individually or collectively as a polaronic entity Kondo-scatter the conduction electrons, leading to the resistivity minimum. Recent electronic structure calculation and spectroscopic investigation attribute the localized moments to oxygen vacancies in the STO near the interface, which trigger Ti (d$^0$) to Ti (d$^1$) conversion and the formation of localized moments \cite{Popovic}. However, it should be noticed that the features seen in R$_\Box$(T) alone are not sufficient to establish unambiguously the presence of Kondo scattering, as such features can also arise from quantum correction to conductivity of a 2D electron system \cite{Bergmann, Lee&Ramakrishnan}. One clear attribute of Kondo scattering is a giant thermoelectric power (S) and electronic specific heat \cite{Bader}, which emanates from a sharply varying energy derivation of the density of states at the Fermi sea due to the formation of Kondo resonance. The Kondo contribution to thermopower as calculated by J. Kondo \cite{Kondo2} is independent of temperature and magnetic impurity concentration in the paramagnetic regime. At low temperature, where the energy scale K$_B$T becomes less than the Zeeman splitting of impurity spins, the S(T) tends to zero as 1/T. A generalized form of Kondo contribution to S(T) is;
\begin{eqnarray}
S(T) = \Lambda \left( {{\raise0.7ex\hbox{${k_B }$} \!\mathord{\left/
 {\vphantom {{k_B } e}}\right.\kern-\nulldelimiterspace}
\!\lower0.7ex\hbox{$e$}}} \right){\raise0.7ex\hbox{$T$} \!\mathord{\left/
 {\vphantom {T {\left( {T + T_0 } \right)}}}\right.\kern-\nulldelimiterspace}
\!\lower0.7ex\hbox{${\left( {T + T_0 } \right)}$}}
\end{eqnarray}
where k$_B$/e = 86 $\mu$V/K and  T$_0$ $\sim$ T$_K$, the Kondo temperature below which the impurity spin is completely screened by conduction electrons extended over a radius of $\sim$ hv$_F$/k$_B$T, where v$_F$ is the Fermi velocity. However, experiments on dilute alloys shows a peak in TEP near the Kondo temperature and its temperature dependence can be evaluated using the Nordheim-Gorter rule which weights the contribution of different scattering mechanisms over a wide temperature scale to the total diffusion thermopower \cite{Matusiak}. In Kondo systems, the entropy carried by the spin degree of freedom is expected to be suppressed by an external magnetic field, and therefore, the S(T) should show magnetic field effects in the Kondo regime. Some early measurements on gold alloys show large suppression of thermopower by a magnetic field \cite{Berman, Huntley}. Clearly, measurements of thermopower and its magnetic field dependence in LAO/STO type 2DEG system can provide important clue for the origin of resistivity minimum and magnetic correlations. Here we report measurements of S(T, H) over a broad range of temperature (T) and magnetic field strength (H) in LTO/STO  2DEG system where a $\delta$ uc thick layer of LaCrO$_3$ (LCO) has been introduced at the interface. In agreement with the earlier studies \cite{Shubhankar, Colby, Chambers}, the pure LCO/STO interface remains insulating. However a conducting interface is realized when m uc layers of LTO are deposited over the $\delta$ uc LCO. Both m and $\delta$ have a critical range for metallic interface. We observe a distinct enhancement in S(T) in the temperature regime where a Kondo-like minimum in R$_\Box$(T) is seen. This enhancement is accentuated further by the presence of Cr$^{3+}$ ions at the interface which also make the R$_\Box$(T) minimum deeper. These observations along with the fact that the S(T) is quenched strongly by a magnetic field in the range of temperature where R$_\Box$(T) shows the minimum, support the Kondo scattering picture of electronic transport in these interfaces.

The films are deposited using KrF eximer laser based pulsed laser deposition (PLD) technique on TiO$_2$ terminated STO (001) single crystal substrates at 800$^0$ C in 10$^{-4}$ mbar oxygen pressure as described earlier in detail \cite {Rastogi, Shubhankar}. To realize a TiO$_2$ terminated, defect-free STO (001) surface, the substrates were treated with HF buffer solution followed by annealing at 800$^0$ C for an hour. The laser was fired at a repetition rate of 1 Hz and a fluence of 1.2 J/cm$^2$ on polycrystalline LTO and LCO targets to get a growth rate of 0.01 nm/s. The atomic and chemical states of the interface have been examined using X-ray reflectivity and cross sectional electron microcopy in conjunction with electron energy loss spectroscopy (EELS) \cite{Shubhankar}. Ag/Cr electrodes were deposited by thermal evaporation through a shadow mask. The S(T, H) measurements were performed on three films - pure LTO/STO and, 1 and 5 uc LCO doped LTO/STO heterostructures. A home-made sample holder mounted on the commercial sample puck (as sketched in Fig. 1(b)) of a physical properties measurement system (PPMS - Quantum Design) equipped with 14 tesla ($\mathcal{T}$) superconducting magnet was used for the measurements of S(T, H) \cite{Setup}. The resistivity and thermopower were measured over a broad temperature range from 5 K to 300 K and up to 13 $\mathcal{T}$ field applied in the plane as well as out-of-plane of the interface. The experimental setup has been calibrated by measuring the thermopower of bismuth and YBa$_2$Cu$_3$O$_7$ superconducting films. The vanishing S(T) of superconducting films below T$_C$ and its strong magnetic field dependence \cite{Budhani} are ideal to test the accuracy of the system.

\begin{figure}[h]
\begin{center}
\includegraphics [ trim=0cm 0cm 0cm 0cm, width=8.6cm, angle=0 ]{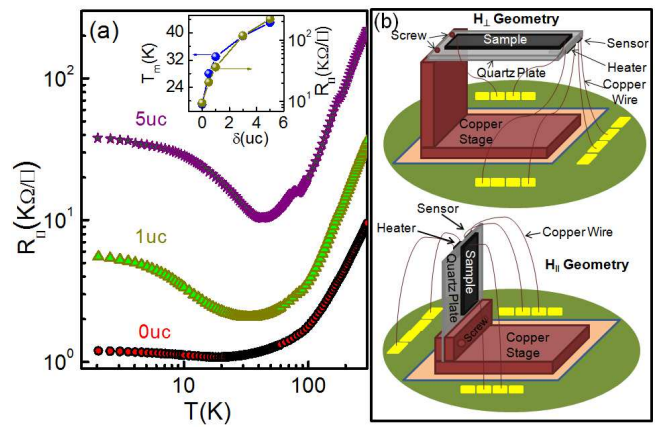}%
\end{center}
\caption{\label{fig1} The temperature dependence of R$_\Box$ for the LTO(25 uc)/LCO ($\delta$-uc)/STO heterostructures where $\delta$ = 0, 1, 5 uc. In the inset of the figure, we have plotted the temperature (T$_m$) at which the minimum in R$_\Box$(T) is seen along with the room temperature R$_\Box$ as a function of LCO layer thickness. (b) Sketch of the platform for measurements of S(T, H) in the perpendicular (H$_\perp$) and parallel (H$_\parallel$) field geometry. The setup is mounted on the DC resistivity puck of PPMS.}
\end{figure}

In Fig. 1 we have shown the R$_\Box$ vs T plots for LTO (25 uc)/LCO ($\delta$-uc)/STO heterostructures; where $\delta$ = 0, 1 and 5 uc. All the samples show a metallic behavior upon lowering the temperature from 300 K. However, at a certain temperature below 100 K, the sheet resistance goes through a minimum followed by a ln T increase and finally saturate at still lower temperatures. The minimum in R$_\Box$(T) shifts towards higher temperature with the increasing doping and the upturn become more prominent. The relevant electronic processes that can produce the observed upturn are: weak localization, 2D electron-electron interaction and Kondo scattering. In our previous work \cite{Shubhankar}, we have established through extensive MR measurements that the Kondo effect combined with a Rashba type spin-orbit scattering process can fully explain the temperature, field and angular dependence of sheet resistance. Here it needs to be mentioned that the Ti$^{3+}$ ions with spin-charge polaronic character contributing to the less dispersed t$_{2g}$ bands \cite{Chang} are on the STO side of the interface. These presumably Kondo scatter electrons in pure LTO/STO system. In the case of $\delta$-doped samples even Cr$^{3+}$ (S = 3/2) can Kondo scatter provided the antiferromagnetic order in LCO is broken close to LCO/STO interface due to non-zero intermixing as suggested by EELS and photoelectron spectroscopy measurements \cite{Shubhankar, Colby, Chambers}. Uncompensated Cr$^{3+}$ spins can also result from incomplete antiferromagnetic order when odd number of LCO unit cells are deposited. In any case, going by the literature on classical Kondo systems, a very dilute concentration of uncompensated spins at the interface may influence transport significantly. In the inset (b) of Fig. 1 we have plotted the temperature T$_m$ at which the minimum in R$_\Box$ is seen along with the value of R$_\Box$ at room temperature as a function of $\delta$-layer thickness. At the higher doping level ($\delta$ $>$ 5 uc) we always see a kink in R$_\Box$ at a temperature slightly higher than T$_m$, whose origin is still under investigation. This feature has been reported earlier as well \cite{Siemons}.

\begin{figure}[h]
\begin{center}
\includegraphics [width=8.8cm]{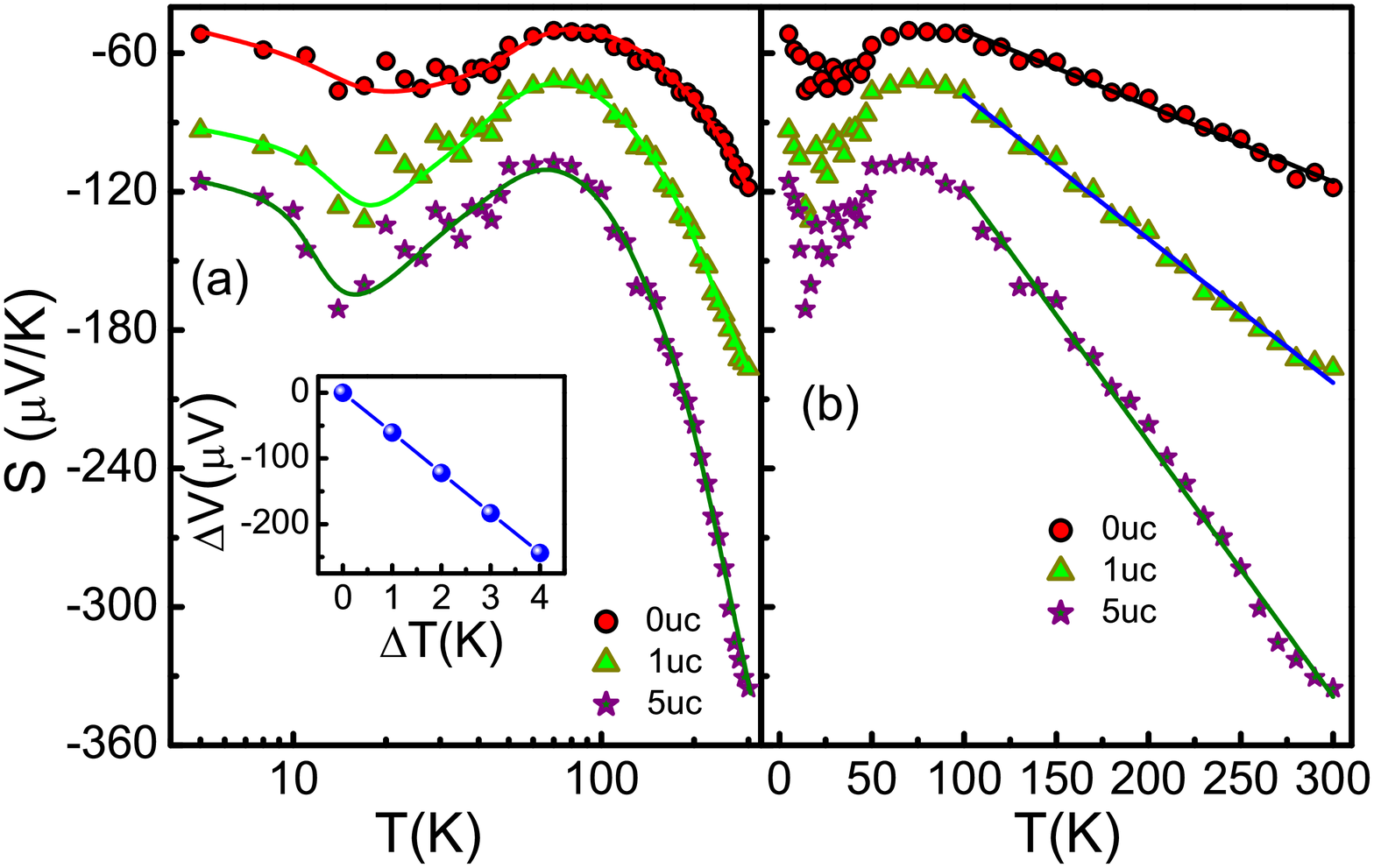}%
\end{center}
\caption{\label{fig2} (a) Zero-field thermopower is shown as a function of logarithmic temperature for $\delta$ = 0, 1 and 5 uc samples. The solid lines are guide to the eye. Inset shows the voltage difference ($\Delta$V) between the ends of the sample as a function of temperature gradient $\Delta$T. (b) the same data of thermopower are plotted in a linear temperature scale to highlight the T-dependence of thermopower at the higher temperatures ($\geq$ 100 K). The solid lines are a fit to Eq. (3).}
\end{figure}

The zero-field thermopower of three samples ($\delta$ = 0, 1 and 5 uc), is shown in Fig. 2(a) in a logarithmic temperature scale to highlight its behavior in the regime where a minimum is seen in the R$_\Box$(T). The S(T) has been calculated from the linear response regime of thermal voltage vs temperature gradient curve as shown in the inset of Fig. 2(a). The negative sign of the voltage indicates electrons as the majority charge carriers in the system, in consistency with the results of Hall measurements \cite{Shubhankar, Biscaras2}. We notice a distinct broad maximum in the magnitude of S(T) over the temperature range of the resistance minimum. Also, there is a striking increase in the overall magnitude of thermopower on $\delta$-doping. At still higher temperatures, the S(T) first drops and then increases monotonically with temperature. To better understand the temperature dependence of thermopower on approach to ambient temperature, we have replotted these data on a linear temperature scale in Fig. 2(b). A linear temperature dependence is unmistakable in the temperature range of $\sim$ 100 K to 300 K.

To understand these data in a semi-quantitative manner, we start with the Mott formula for diffusion thermopower of disordered metal \cite{Mott&Karavolas&Sankeshwar};
\begin{eqnarray}
S^d  = \frac{{\pi ^2 }}{3}\left( {\frac{{k_B^2 T}}{e}} \right)\left\{ {\frac{1}{n}\frac{{\partial n(\varepsilon )}}{{\partial \varepsilon }} + \frac{1}{\mu }\frac{{\partial \mu (\varepsilon )}}{{\partial (\varepsilon )}}} \right\}_{\varepsilon _F }
\end{eqnarray}
where n and $\mu$ are the carrier density and their mobility respectively. The energy derivative of the density of states (N($\varepsilon$)) near Fermi energy ($\varepsilon$$_F$) depends on the dimensionality of the system; as for a low dimensional material N($\varepsilon$) has singularities at the specific values of energy. The N($\varepsilon$) distribution is also affected by electron - electron interaction effects at low temperatures and the Kondo - type resonance introduced by localized magnetic impurities. While the energy dependence of mobility can be ignored in a highly disordered limit where the mean free path is of the order of interatomic distance, in cleaner samples one needs to consider the second term of Eq. (2) as well.

For a 2DEG in the absence of Kondo resonance, the diffusion thermopower can be expressed as \cite{Mott&Karavolas&Sankeshwar};
\begin{eqnarray}
S^d  = \frac{{\pi ^2 k_B^2 }}{{3e}}\frac{T}{{\varepsilon _F }}(p - 1)
\end{eqnarray}
where the parameter p($\varepsilon$) takes into account the energy dependence of the electron scattering time due to all scattering mechanism using Matthissen's rule \cite{Mott&Karavolas&Sankeshwar}. In the case of a 2DEG confined in the z-direction by interfaces, the interface scattering contribution is added to p($\varepsilon$). The solid lines in Fig. 2(b) are fits to Eq. (3). Here the Fermi energy of the three samples has been calculated from the measured carrier concentration. Table I lists the E$_F$ and p($\varepsilon$) parameters extracted from the best fits.
\begin{table}[h]
\begin{center}\caption{The values of sheet carrier density (n$_\Box$), Fermi energy and the parameter p are listed for $\delta$ = 0, 1, 5 uc samples.} \label{TABLE I.}
\begin{tabular} {|c|c|c|c|}
\hline
Doping (uc) & n$_\Box$($\setminus$cm$^2$) & $\varepsilon$$_F$(meV) & p \\
\hline
$\delta$ = 0 & 3 $\times$ 10$^{14}$ & 712.8 & 10.55 \\
\hline
$\delta$ = 1 & 5.3 $\times$ 10$^{13}$  & 126.2 & 4.21 \\
\hline
$\delta$ = 5 & 1.3 $\times$ 10$^{12}$  & 3.1 & 1.14 \\
\hline
\end{tabular}
\end{center}
\end{table}

In a magnetic system, this monotonic temperature dependence expressed by Eq. (2) may be affected by the Kondo contribution to thermopower at low temperatures, which arises from the higher order s-d scattering processes. At T $<$ T$_K$, where T$_K$ $\sim$ T$_F$ exp(-1/JN) and J is the exchange constant between conduction electron spin $\vec S_e$ and the impurity spin $\vec S_i$, the Kondo contribution to thermopower is given by Eq. (1). It leads to a maximum in diffusion thermopower in the vicinity of T$_K$. However, before we attribute the enhancement seen in Fig. 2(a) to Kondo scattering, it is worthwhile to address the contribution of phonons to thermopower. An enhancement in thermopower is also expected from the momentum transferred to charge carriers by phonons drifting along the temperature gradient. The phonon drag thermopower can be expressed as S$^{ph}$ $\sim$ BT$^q$, where B is phonon mean free path and the exponent q takes positive values \cite{Fletcher}. A relatively simple calculation of phonon drag thermopower under the assumption that phonon-phonon collisions are few yields S$^{ph}$ = C/3ne, where C is the lattice specific heat per unit volume and n the carrier concentration. The S$^{ph}$ is expected to peak at the temperature close to one fifth of the Debye temperature (T$_D$ $\approx$ 400 K) for pure STO. Since the 2D gas is in the STO side of the interface, we expect the number density of phonons to be decided by the STO substrate.

In order to establish whether the peak in S(T) is due to phonon drag or Kondo scattering, we have measured the thermopower as a function of magnetic field in the temperature range of the peak. It is expected that in the regime of Kondo cloud formation by conduction electrons to screen the non-interacting impurity spins, electrons while going from one cloud to the next having oppositely oriented impurity spin will transport spin entropy and hence contribute to thermopower. An external magnetic field tends to promotes ordering of impurity spins and hence is expected to reduce the S(T, H) \cite{Ong, Choi}.

\begin{figure}[h]
\begin{center}
\includegraphics [width=8.4cm]{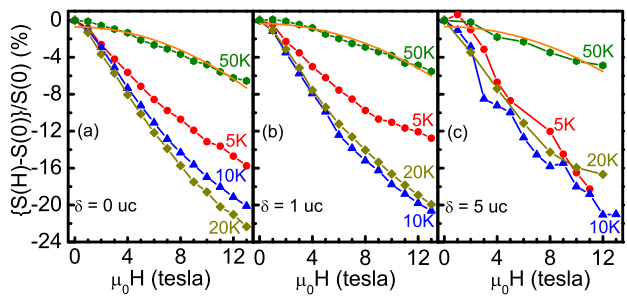}%
\end{center}
\caption{\label{fig3} (a)-(c) show the magnetic field dependence of thermopower for $\delta$ = 0, 1 and 5 uc samples at various temperatures. Maximum drop in S is observed in the vicinity of the temperature where a minimum in R$_\Box$ is seen. The thermopower shows negligible field dependnce above 100 K (not shown in figure). The solid orange curve for 50 K data in all the figures is a fit using A + BH$^2$.}
\end{figure}

In Fig. 3 (a, b, c) we present the magnetothermopower defined as $\frac{{S(H) - S(0)}}{{S(0)}}$ in percentage for $\delta$ = 0, 1 and 5 uc samples at several temperatures. The external magnetic field in this case was applied perpendicular to the plane of the sample (H$_\perp$). A comparison of these data and the behavior of R$_\Box$(T) makes it clear that $\mid$S(H)$\mid$ is strongly suppressed in the vicinity of the temperature where R$_\Box$(T) goes through a minimum. The reduction in thermopower is as high as 20\% in a 13 $\mathcal{T}$ field. The shape of the $\Delta$S(H) vs H curves also indicates that the change is steeper at intermediate fields followed by a tendency for saturation at higher fields, which has been seen earlier also in Kondo systems and magnetic alloys \cite{Berman, Huntley}. This is indicative of a field induced ordering of spins in the system and consequent suppression of spin entropy transport.

Weiner and Beal-Monod have given a rigorous calculation of thermopower of Kondo alloys in the frame work of s-d scattering \cite{Weiner}. The relevant energy scales in the problem are k$_B$T and g$\mu$$_B$H. In the regime of low field (g$\mu$$_B$H / k$_B$T $<$ 1), they predict $\Delta$S(H)/S(0) $\sim$ M$^2$ $\sim$ H$^2$, while at high field (g$\mu$$_B$H / k$_B$T $>$ 1), $\Delta$S(H)/S(0) $\sim$ 1/H.

\begin{figure}[h]
\begin{center}
\includegraphics [width=8.4cm]{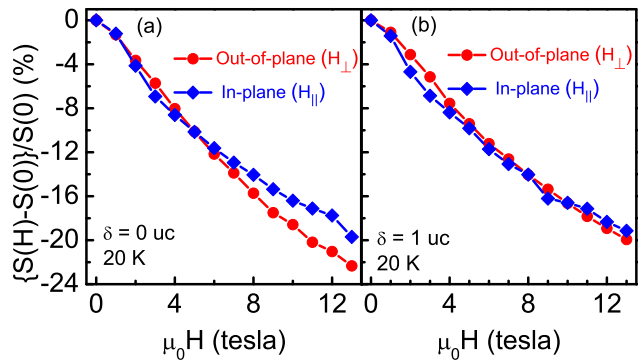}%
\end{center}
\caption{\label{fig4} (a) and (b) show a comparison between the H$_\perp$ and H$_\parallel$ magnetothermopower at 20 K for $\delta$ = 0 and 1 uc samples respectively. The field direction independence of S(H, T) supports the Kondo picture of electronic transport in such heterostructures.}
\end{figure}

Although in our case the condition g$\mu_B$H $<$ K$_B$T is satisfied when H $<$ 3.7, 7.4 and 14.9 $\mathcal{T}$ at T = 5, 10 and 20 K respectively, the saturation tendency of impurity spin magnetization will certainly lead a less steeper change in S at the higher fields, which is seen in Fig. 3(a-c).

It should be pointed out that the magnetothermopower of a Kondo system is expected to be isotropic in field direction. To verify this aspect of the S(H), we have also measured the thermopower in a configuration where the magnetic field is in the plane of the interface (H$_\parallel$) and along the direction of temperature gradient. The results of these measurements for $\delta$ = 0 and 1 uc samples at 20 K are shown in Fig. 4(a) and (b) respectively. Here we also show, for comparison, the S(H) data taken in H$_\perp$ configuration. Interestingly, the behavior of S(H) for H$_\perp$ and H$_\parallel$ is similar. While this observation is consistent with the Kondo picture, it needs to be pointed out that the magnetoresistance of these interfaces show strong anisotropy in H$_\perp$ and H$_\parallel$ configuration \cite{Shubhankar}. The MR for H$_\perp$ is positive and goes as $\approx$ H$^2$, suggesting enhanced classical scattering of conduction electrons as they go into the cyclotron orbit. The MR for H$_\parallel$ is comparatively much smaller and negative in sign, in consistency with the Kondo scattering mechanism. While this higher order s-d scattering process will also contribute to the MR in H$_\perp$ geometry, its magnitude is much smaller than the MR from orbital scattering. The relatively isotropic magneto-thermopower seen in these systems suggest that the energy derivative of mobility (Eq. (1)) makes negligible contribution to thermopower as compared to the term which involves the density of states, and the influence of magnetic field on DOS is nontrivial.

In summary, we have carried out detailed measurements of thermoelectric power and its magnetic field dependence in LTO/STO 2DEG samples where the interface has been selectively modified by inserting a few monolayers of LCO. The S(T) in these systems is large and negative, with a distinct enhancement in the temperature range of 10 K $\sim$ 50 K, where the R$_\Box$(T) goes through a minimum. The thermopower in this range of temperature is suppressed significantly ($\leq$ 20\%) by a moderate magnetic field ($\leq$ 13 $\mathcal{T}$) applied either normal or parallel to the plane of the interface. These features of magnetothermopower strongly suggest Kondo scattering of charge carriers by localized magnetic moments in this system. It is also established that the S(T) at higher temperatures follows the theory for a two-dimensional conductor.

SD and AR thank Indian Institute of technology, Kanpur and CSIR-India for financial help. RCB acknowledges J. C. Bose national fellowship of the Department of Science and Technology.

\end{document}